\documentclass[pageno]{jpaper}

\usepackage{enumitem}
\usepackage{xcolor}
\usepackage{cite}
\usepackage{amsmath}
\usepackage{float}
\usepackage{caption}
\usepackage{graphicx}
\usepackage[linesnumbered,ruled,vlined]{algorithm2e} 
\usepackage{listings}
\usepackage{calc}  
\usepackage{tabu}
\usepackage{mathtools}
\usepackage{tikz}
\usepackage{makecell}
\usetikzlibrary{arrows.meta}
\usepackage{pgfplots}
\usepackage{multirow}
\usepackage{flushend}
\usepackage{booktabs}
\usepackage{hyperref}
\usepackage[sort, numbers]{natbib}
\usepackage{amssymb}

\definecolor{codegreen}{rgb}{0,0.6,0}
\definecolor{codegray}{rgb}{0.5,0.5,0.5}
\definecolor{codepurple}{rgb}{0.58,0,0.82}
\definecolor{backcolour}{rgb}{0.95,0.95,0.92}
\definecolor{jcredl}{HTML}{fb9a99}
\definecolor{jcgreenl}{HTML}{b2df8a}
\definecolor{jcbluel}{HTML}{a6cee3}
\definecolor{jcorangel}{HTML}{fdbf6f}
\definecolor{jcpurplel}{HTML}{cab2d6}
\definecolor{jcred}{HTML}{e31a1c}
\definecolor{jcbrown}{HTML}{b15928}
\definecolor{jcgreen}{HTML}{33a02c}
\definecolor{jcblue}{HTML}{1f78b4}
\definecolor{jcorange}{HTML}{ff7f00}
\definecolor{jcpurple}{HTML}{6a3d9a}

\lstdefinestyle{mystyle}{
  frame=tblr,
  commentstyle=\color{codegreen},
  keywordstyle=\color{codepurple},
  basicstyle=\footnotesize\ttfamily,
  breakatwhitespace=false,         
  breaklines=true,                 
  captionpos=b,                    
  keepspaces=true,                 
  numbers=left,                    
  numbersep=5pt,                  
  showspaces=false,                
  showstringspaces=false,
  showtabs=false,                  
  tabsize=2,
  escapeinside={(*@}{@*)},
}
\lstdefinelanguage{mlir}{%
  language     = C,
  morekeywords = {affine, for, arith, muli, addi, store, load, step},
}
\lstdefinelanguage{seerlang}{%
  language     = C,
  morekeywords = {block, seq, affine, for, arith, muli, addi, store, load, step},
}

\newcommand*\jc[1]{\textcolor{jcorange}{\bf #1}}

\newcommand*\best[1]{\textcolor{jcgreen}{\bf #1}}
\usepackage{soul}

\newcommand \egg{\texttt{egg}}
\begin{document}

\title{SEER: Super-Optimization Explorer for HLS using E-graph Rewriting with MLIR}

\date{}
\maketitle

\thispagestyle{empty}

\begin{abstract}
High-level synthesis (HLS) is a process that automatically translates a software program in a high-level language into a low-level hardware description. However, the hardware designs produced by HLS tools still suffer from a significant performance gap compared to manual implementations. This is because the input HLS programs must still be written using hardware design principles.

Existing techniques either leave the program source unchanged or perform a fixed sequence of source transformation passes, potentially missing opportunities to find the optimal design. We propose a super-optimization approach for HLS that automatically rewrites an arbitrary software program into efficient HLS code that can be used to generate an optimized hardware design. We developed a toolflow named SEER, based on the e-graph data structure, to efficiently explore equivalent implementations of a program at scale. SEER provides an extensible framework, orchestrating existing software compiler passes and hardware synthesis optimizers. 

Our work is the first attempt to exploit e-graph rewriting for large software compiler frameworks, such as MLIR. Across a set of open-source benchmarks, we show that SEER achieves up to 38$\times$ the performance within 1.4$\times$ the area of the original program. Via an Intel-provided case study, SEER demonstrates the potential to outperform manually optimized designs produced by hardware experts.
\end{abstract}

%%%%%%%%%%%%%%%%%%%%%%%%%%%%%%%%%%%%%%%%%%%%%%%%%%%%%
% INTRODUCTION
%%%%%%%%%%%%%%%%%%%%%%%%%%%%%%%%%%%%%%%%%%%%%%%%%%%%%
\section{Introduction}
% \jc{TODO: we have mentioned many examples here but have we made it clear that it is actually a general problem?}
High-level synthesis (HLS) is a process that automatically translates a software program in a high-level language such as C/C++ into a hardware description in a low-level language such as Verilog/VHDL. This allows software engineers without any hardware background to customize their hardware accelerators. Today, HLS tools have been widely used and actively developed in both academia and industry, for example, Dynamatic~\cite{josipovic_fpga2018} from EPFL, Bambu~\cite{castellana_hcs2014} from the Politecnico di Milano, Stratus HLS~\cite{stratus_hls} from Cadence, Catapult HLS~\cite{catapult_hls} from Siemens, Intel HLS~\cite{intel_hls} from Intel and Vitis HLS~\cite{vitis_hls} from AMD Xilinx.

% \jc{? => Phase-ordering => HLS worse => control path/data path/gate level => e-graph} \gc{Multi-level phase ordering problem.}

\begin{figure}
\centering
\begin{minipage}[b][][b]{.45\textwidth}
\begin{tabular}{r}
\includegraphics[scale=0.6]{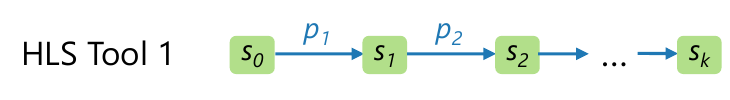} \\
\includegraphics[scale=0.6]{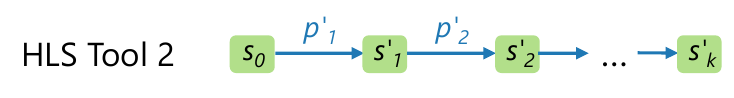} \\
\includegraphics[scale=0.6]{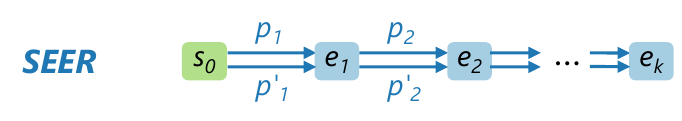}
\end{tabular}
\end{minipage}
\caption{$p_i$ denotes an optimization pass, and $s_i$ denotes a representation of a program. HLS Tools 1 and 2 take the same input program $s_0$ and apply different sequences of optimization passes, $p_i$ and $p'_i$ respectively. The transformed programs $s_k$ and $s'_k$ may result in different hardware designs because of the difference in pass sequences. SEER efficiently explores all these possibilities in parallel using e-graphs, $e_i$. 
% The e-graph retains a complete history such that $\forall j\leq i\;s_j, s_j'\in e_i$.
}
\label{fig:problem_formulation}
\end{figure}

Still, it remains the case that automatically synthesizing efficient hardware designs from arbitrary high-level software programs is challenging. A major reason is that each HLS tool only applies a fixed sequence of general source transformations for all input programs, as shown in Figure~\ref{fig:problem_formulation}. This significantly restricts the optimization space for a particular program. This is known as the {\bf {\em `phase-ordering problem'}} in compilers~\cite{leverett1980overview}.

The phase-ordering problem for HLS tools is more challenging for two reasons. First, an HLS tool contains optimizations at {\bf {\em different granularities}}, such as higher-level control path optimizations and lower-level data path optimizations. These optimizations may interfere, resulting in a larger, more complex space of optimization orderings than in software compilers. Second, evaluating {\bf {\em hardware metrics}} from an input software program is challenging in existing frameworks. This means that the optimizer needs to repeatedly call the downstream synthesis tool to evaluate which source representation is efficient when mapping into hardware.

Existing works on HLS source rewriting build an optimization sequence based on heuristics. This misses opportunities to perform program-specific optimizations for a given input program, potentially making the optimal hardware design unreachable. In practice, significant manual effort is spent on rewriting the program source for HLS tools to resolve the problem above. Both Stratus HLS~\cite{stratus_hls} and Vitis HLS~\cite{vitis_hls}, provide recommended coding styles in their user manual to restrict users to a subset of C programs for better performance. A designer must write the HLS program following these guidelines and using hardware design principles in order to produce efficient hardware.

\begin{figure*}
\centering
\begin{tabular}{m{0.25\textwidth} m{0.04\textwidth} m{0.25\textwidth} m{0.04\textwidth} m{0.25\textwidth}}
\begin{lstlisting}[language=C, escapeinside={(*@}{@*)}, caption=Baseline code, label=list:1, linewidth = 0.25\textwidth]
int x[200], y[200];

loop_1:
for (int i=0; i<100; i++)
  x[i+1] = f(x[i]);

loop_2:
for (int i=0; i<100; i++)
  y[i] = g(y[i]);

loop_3:
for (int i=0; i<100; i++)
  x[i+2] = h(x[i]);
\end{lstlisting}
&
\includegraphics[width=0.02\textwidth]{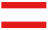}
&
\begin{lstlisting}[language=C, escapeinside={(*@}{@*)}, caption=Transform 1, label=list:2, linewidth = 0.25\textwidth]
int x[200], y[200];

// Fuse loop_1 and loop_2
loop_1_2:
for (int i=0; i<100; i++) {
  x[i+1] = f(x[i]);
  y[i]   = g(y[i]);
}


loop_3:
for (int i=0; i<100; i++)
  x[i+2] = h(x[i]);
\end{lstlisting}
&
\includegraphics[width=0.02\textwidth]{imgs/equal.pdf}
&
\begin{lstlisting}[language=C, escapeinside={(*@}{@*)}, caption=Transform 2, label=list:3, linewidth = 0.25\textwidth]
int x[200], y[200];

loop_1:
for (int i=0; i<100; i++)
  x[i+1] = f(x[i]);


// Fuse loop_2 and loop_3
loop_2_3:
for (int i=0; i<100; i++) {
  y[i]   = g(y[i]);
  x[i+2] = h(x[i]);
}
\end{lstlisting}
\end{tabular}
\caption{A motivating example of loop fusion. {\tt loop\_1} and {\tt loop\_3} cannot be fused because of the memory dependence on array {\tt x}. It is challenging to determine which representation is better, fusing {\tt loop\_1} and {\tt loop\_2} or fusing {\tt loop\_2} and {\tt loop\_3}?}
\label{fig:motivating_example}
\end{figure*}

In order to tackle the problems above, our work aims to solve the following challenges:
\begin{description}[leftmargin=!]
    \item[1) Efficiency: ] How should one efficiently explore the vast space of possible optimization sequences?
    \item[2) Hardware awareness: ] How should one pick a sequence of optimizations that can be mapped into efficient hardware based on the program source?
\end{description}
We propose an approach named SEER (\textbf{S}uper-optimization \textbf{E}xplorer using \textbf{E}-graph \textbf{R}ewriting) to resolve the challenges above. \emph{Given a software program, SEER automatically determines a sequence of optimizations for efficient hardware synthesis}. SEER is the first approach to HLS `super-optimization', since it explores different source-level optimization orderings in parallel, then customizes the sequence to the input program.

SEER enables super-optimization using an efficient data structure, known as an e(quivalence)-graph~\cite{chao1978chromatic}, which preserves a set of program representations to resolve the phase-ordering problem. As shown in Figure~\ref{fig:problem_formulation}, SEER can explore alternative optimization sequences in a single e-graph at the same time. Our main research contributions include: 
\begin{itemize}
    \item a technique to determine an optimization order for efficient hardware synthesis by exploring equivalent representations of a program in an e-graph;
    \item an orchestration technique that explores an e-graph with existing optimization passes from large software frameworks, such as MLIR~\cite{lattner2021mlir}, and hardware synthesis optimizers, such as ROVER~\cite{coward2022rover}, to explore rewriting at scale;
    \item a hardware-aware evaluation model at the source level to evaluate the quality of hardware synthesized from a representation of a software program; and
    \item over a set of benchmarks, SEER achieves up to 38$\times$ the performance within 1.4$\times$ the area of the original program, and demonstrates the potential to outperform manually optimized designs by hardware experts.
\end{itemize}

The rest of the paper is organized as follows. Section~\ref{sec:motivating_example} presents a motivating example to illustrate the challenge in automated source rewriting for HLS. Section~\ref{sec:background} provides the necessary background and related work. Section~\ref{sec:methodology} explains the theoretical details of our work. Section~\ref{sec:experiments} evaluates the effectiveness of our work by comparing it with vanilla Stratus HLS and manual design by experts. 

%%%%%%%%%%%%%%%%%%%%%%%%%%%%%%%%%%%%%%%%%%%%%%%%%%%%%
% MOTIVATING EXAMPLE
%%%%%%%%%%%%%%%%%%%%%%%%%%%%%%%%%%%%%%%%%%%%%%%%%%%%%
\section{Motivating Example}
\label{sec:motivating_example}

Using an example, we present the challenge to conventional, fixed pass-order HLS flows and how our approach can overcome these challenges.
Listing~\ref{list:1} presents a program with three sequential loops. 
Loop fusion is an optimization technique that combines multiple sequential loops into a single loop. In HLS, loop fusion avoids the area overhead of the loop control logic for separate loop instances and could exploit more data parallelism in the fused loop body. However, the throughput of a loop is restricted by the slowest data path in the loop body. Loop fusion can exhibit an area-performance tradeoff.

A pre-condition for loop fusion being valid is that the sequential loops must have no data dependence. For the example in Listing~\ref{list:1}, {\tt loop\_1} and {\tt loop\_3} access array {\tt x} at overlapping indices, preventing loop fusion. However, {\tt loop\_2} has no data dependence with either {\tt loop\_1} or {\tt loop\_3}, since it only accesses array {\tt y}. This means that we can safely fuse {\tt loop\_1} and {\tt loop\_2} (Listing~\ref{list:2}) or fuse {\tt loop\_2} and {\tt loop\_3} (Listing~\ref{list:3}). 
%All three representations in Figure~\ref{fig:motivating_example} are functionally equivalent but may exhibit different hardware performance, area and power, when synthesized. 
Note that the user or automated tool must choose between these fusion passes, since Listing~\ref{list:2} is not reachable from Listing~\ref{list:3}, and vice versa.

\begin{table}
\centering
\caption{The performance of hardware generated from the representations in Figure~\ref{fig:motivating_example} depends on the operation latencies. These could be affected by other transformation passes and are not evaluated in the existing flow. The best performance in each case is \best{highlighted}.}
\label{tab:motivating_example}
\resizebox{0.48\textwidth}{!}{%
\begin{tabular}{crrrrrr}
\toprule
 & \multicolumn{1}{c}{\tt f} & \multicolumn{1}{c}{\tt g} & \multicolumn{1}{c}{\tt h} & \multicolumn{1}{c}{Listing~\ref{list:1}} & \multicolumn{1}{c}{Listing~\ref{list:2}} & \multicolumn{1}{c}{Listing~\ref{list:3}} \\
 \midrule
Case 1 & 10 & 100 & 1 & 1308 & \best{1196} & 1205 \\
Case 2 & 1 & 100 & 10 & 813 & 710 & \best{701} \\
\bottomrule
\end{tabular}}
\end{table}

Without evaluating downstream hardware optimization passes, it is difficult to determine whether Listing~\ref{list:2} or~\ref{list:3} will generate better hardware. Table~\ref{tab:motivating_example} shows the performance of the hardware generated from these representations for different latencies of the functions {\tt f} and {\tt h}. Such latency information is unpredictable at the source rewriting stage because later passes might alter these functions. This correlation could make a locally sub-optimal transformation globally optimal. SEER models hardware scheduling information in software and efficiently explores transformations of these representations in an e-graph instead of manipulating a single representation. 

\subsection*{Problem Formalization}

A key novelty of our work is that SEER explores the correlation among transformation passes, which opens up a larger design space. Let $P$ be a set of available transformation passes, $P^{\mathbb{N}}$ be all possible sequences of the elements in $P$ and let $R$ be the set of functionally equivalent representations of a given program. As shown in Figure~\ref{fig:problem_formulation}, an HLS tool uses a fixed sequence of passes $t = (p_0, p_1, ..., p_k)\in P^{\mathbb{N}}$. The transformation steps in $t$ result in a set of representations $R'$, where $R' \subseteq R$. SEER searches the space of pass sequences $P^{\mathbb{N}}$ and extracts a customized $t' \in P^{\mathbb{N}}$ for each input program. SEER can explore a potentially larger set of representations $R''$, where $R' \subseteq R'' \subseteq R$. This is because SEER searches for $t'$ by exploring $P^{\mathbb{N}}$ in parallel. In the rest of the paper, we show how to construct $R''$ using an e-graph and how to determine $t'$ for mapping an arbitrary program to efficient hardware. 

%%%%%%%%%%%%%%%%%%%%%%%%%%%%%%%%%%%%%%%%%%%%%%%%%%%%%
% BACKGROUND
%%%%%%%%%%%%%%%%%%%%%%%%%%%%%%%%%%%%%%%%%%%%%%%%%%%%%
\section{Background}
\label{sec:background}

% \jc{Comment the following out if we need space}
% In this section, we first review the phase-ordering problem in HLS tools. Then we introduce background about e-graph and related works. Finally, we reviewed related works to HLS in Multi-Level Intermediate Representation (MLIR). 

\begin{figure}
    \centering
    \includegraphics[width=0.5\textwidth]{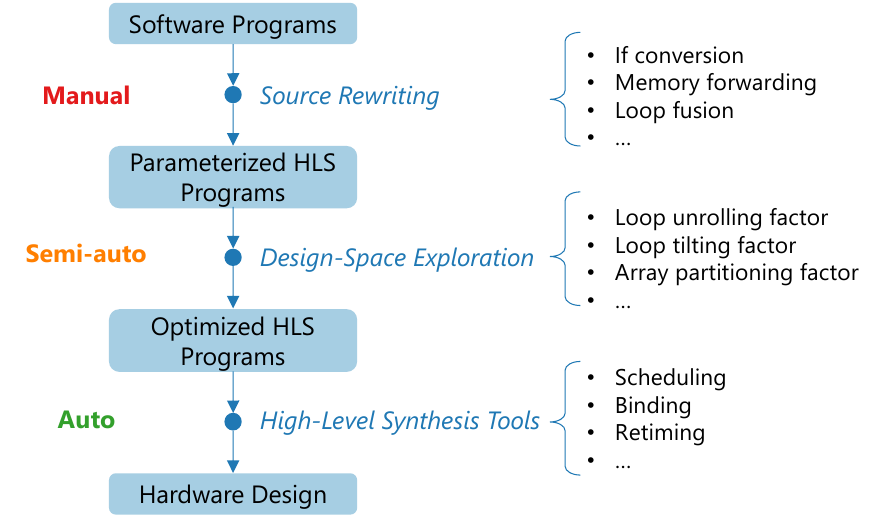}
    \caption{HLS development flow for hardware production. The right side provides examples of optimizations for each step. SEER aims to solve the challenge in efficient source rewriting for arbitrary programs (shown as {\bf \color{jcred}{Manual}}) for better hardware performance.}
    \label{fig:old_hls_flow}
\end{figure}

%%%%%%%%%%%%%%%%%%%%%%%%%%%%%%%%%%%%%%%%%%%%%%%%%%%%%
% PHASE ORDERING and HLS
%%%%%%%%%%%%%%%%%%%%%%%%%%%%%%%%%%%%%%%%%%%%%%%%%%%%%
\subsection{Phase-Ordering Challenges in High-Level Synthesis}
HLS tools automatically map a high-level software program into a custom hardware design in a low-level hardware description, {\em e.g.} Verilog. A production HLS development flow comprises three steps, as shown in Figure~\ref{fig:old_hls_flow}. First, a high-level specification of an algorithm is \emph{manually} rewritten following the recommended coding guidelines producing code that is amenable to optimization by the HLS tool. Second, the rewritten HLS program usually contains design constraints expressed via inline directives or pragmas to exploit hardware parallelism and resource sharing. The process of exploring these constraints is known as design-space exploration~(DSE)~\cite{schafer2019high} and is already semi-automated~\cite{krishnan2006genetic, liu2013learning, ferretti2020leveraging}. Finally, the optimized design constraints are sent with the HLS program to the HLS tool, which synthesizes a hardware design. The HLS tool automatically performs low-level hardware optimizations, such as hardware scheduling and binding, which maps the start times of operations into clock cycles with efficient hardware resource sharing~\cite{canis2014modulo, zhang2013sdc,hara2012selective, krishnapriya2018high}. The HLS tool also performs register retiming to achieve a high clock frequency~\cite{hara2012selective, krishnapriya2018high}.

% The HLS tool performs four hardware-specific functions. It determines an execution schedule for each operation, implements resource sharing, inserts registers in the design and pipelines loops to maximize performance.

The phase-ordering problem refers to the challenge of determining the optimal order of optimization passes at compile time. It is challenging due to the destructive interaction of optimization passes, as discussed in Section~\ref{sec:motivating_example}. Existing works address the phase-ordering problem for compilers using two approaches. First, there are works that use machine learning~\cite{kulkarni2012mitigating, ashouri2017micomp, ashouri2018survey, huang2019autophase,neto2022flowtune} for inferring a productive sequence of optimization steps. These approaches only work for domain-specific programs, while our approach works for arbitrary programs. Second, there are works that use heuristics-based or iterative approaches for efficient searching for the optimization steps~\cite{nobre2016graph, kulkarni2009practical, zhang2022heterogen}. The intermediate traces during the iterations are not efficiently preserved, while our work carries it in the e-graph during the exploration. All these approaches only target software optimization, while our approach targets hardware optimization. 

In HLS, the phase-order problem is more complex, because the benefit of software transformation passes, such as loop fusion and if conversion, can only be evaluated by analyzing the generated hardware. In this work, SEER orchestrates software transformations for hardware optimization using hardware modeling. To the best of our knowledge, SEER is the first attempt to resolve the general phase-ordering problem in HLS. 
% \jc{This needs to be cut...}

%%%%%%%%%%%%%%%%%%%%%%%%%%%%%%%%%%%%%%%%%%%%%%%%%%%%%
% E-GRAPHS
%%%%%%%%%%%%%%%%%%%%%%%%%%%%%%%%%%%%%%%%%%%%%%%%%%%%%
\subsection{E-graph Representation}

\begin{figure}
\centering
\begin{tabular}{b{0.2\columnwidth} b{0.05\columnwidth} b{0.2\columnwidth} b{0.05\columnwidth} b{0.2\columnwidth}}
\makecell[c]{\tt (x<<1)+x}
&
\makecell[c]{$\rightarrow$}
&
\makecell[c]{\tt (x*2)+x}
&
\makecell[c]{$\rightarrow$}
&
\makecell[c]{\tt x*3}
\\
\\
% \multicolumn{5}{c}{\includegraphics[scale=.8]{imgs/green_down_arrow.pdf}} \\\\
\multicolumn{5}{c}{\includegraphics[scale=.8]{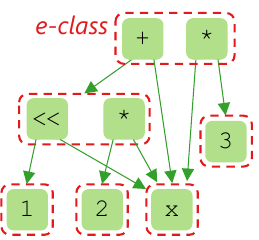}} \\
\end{tabular}
\caption{An e-graph grown from two rewriting steps to represent three equivalent expressions. Each green node is an e-node, and each red box is an e-class. Edges connect e-nodes to child e-classes. 
%\jc{I added the explanation to the annotation. Is that correct?}
}
\label{fig:egraph_example}
\end{figure}

\begin{figure*}
    \centering
    \includegraphics[width=\textwidth]{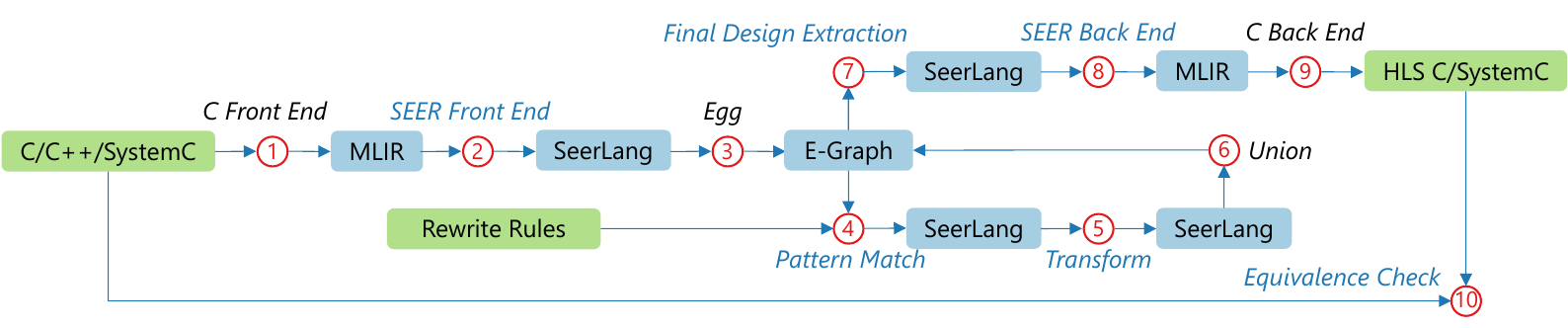}
    \caption{An overview of the SEER toolflow. Our contributions are {\em \color{jcblue}{highlighted}}. 
    }
    \label{fig:seer_overview}
\end{figure*}

An e(quivalence)-graph, is a data structure used to represent a set of equivalent expressions~\cite{Nelson1980TechniquesVerification,willsey2021egg, tate2009equality} as shown in Figure~\ref{fig:egraph_example}. The e-graph organizes functionally equivalent expressions into equivalence classes, known as e-classes, drawn as red boxes in the figure. Nodes in an e-graph, known as e-nodes, represent either values or operators with edges connecting e-class children, illustrated as green nodes in the figure. E-classes are represented as groups of e-nodes. The e-graph is grown via constructive rewriting, meaning that the left-hand side of the rewrite is retained in the data structure. A minimal cost expression is typically extracted from the e-graph, based on a user-defined cost model. 

A main benefit of the e-graph data structure is that it efficiently represents equivalent expressions by sharing and reusing common sub-expressions, such as sharing {\tt x} in Figure~\ref{fig:egraph_example}. Operator e-nodes have edges connected to child e-classes. This captures the intuition that, for a given sub-expression, we can choose from a set of equivalent sub-expressions. The reduced redundancy in the e-graph enables more efficient analysis and optimization of a program.

E-graphs can be found in modern SMT solvers, such as Z3~\cite{DeMoura2008Z3:Solver,de2007efficient}. The recently developed \egg~library~\cite{willsey2021egg}, provides an extensible e-graph implementation that has fueled a new wave of e-graph research. Since its release, e-graphs have been applied to hardware design automation~\cite{coward2022rover,ustun2022impress}, numerical stability improvement~\cite{panchekha2015automatically}, compiler design~\cite{tate2009equality} and much more~\cite{nandi2021rewrite,vanhattum2021vectorization,smith2021pure}. One relevant work~\cite{coward2023combining}, identified improvements to program analysis capabilities by using the e-graph representation. We describe how SEER exploits this in Section~\ref{sec:analysis_capability_enabling}. 
SEER is the first approach to apply e-graphs to program optimization using compiler frameworks such as MLIR.

In this work, we incorporate and extend an existing \egg~based data path optimization engine, named ROVER~\cite{coward2022rover,coward2023automating}. ROVER takes a combinational hardware design and optimizes the data paths using optimizations for circuit area minimization. The existing ROVER implementation leaves the control path untouched, such as loops. SEER generalizes ROVER to a higher-level software abstraction for HLS tools and combines it with control path optimizations for pipelined designs.
% \jc{I think we should explain what ROVER is and why we add ROVER into SEER? I feel we should give as much introduction as MLIR...}

%%%%%%%%%%%%%%%%%%%%%%%%%%%%%%%%%%%%%%%%%%%%%%%%%%%%%
% MLIR
%%%%%%%%%%%%%%%%%%%%%%%%%%%%%%%%%%%%%%%%%%%%%%%%%%%%%
\subsection{Multi-Level Intermediate Representation}
\label{sec:mlir_background}

Multi-Level Intermediate Representation (MLIR)~\cite{lattner2021mlir} is a compiler infrastructure framework developed within the Low-Level Virtual Machine (LLVM) project~\cite{lattner2004llvm}. It aims to address the challenges of representing and optimizing programs at different levels of abstraction. Dialects can be seen as a namespace for operations, types and attributes modelling specific abstractions (i.e. control flow or affine loops). Primarily, SEER uses the {\tt affine} and {\tt scf} dialects. The {\tt affine} dialect provides a program abstraction for affine operations, and the {\tt scf} dialect provides a program abstraction for structured control flows. MLIR offers a comprehensive set of transformation and analysis passes that can be directly reused and explored in SEER.

CIRCT~\cite{circt} is an MLIR-based hardware compiler framework under LLVM, which lowers MLIR to RTL code as an open-sourced HLS tool. Xu~\textit{et al.} propose a specific MLIR dialect named HECTOR for hardware synthesis~\cite{xu2022iccad}, which can be translated into RTL code. There are also source transformation tools that transform MLIR into optimized HLS code in C~\cite{yehpca2022scalehls, lai2019fpga} or LLVM IR~\cite{agostini2022cf, zhao2022fpl}. Both these works and SEER have an end-to-end HLS flow in MLIR. Prior work suffers from the phase-ordering problem illustrated in Figure~\ref{fig:problem_formulation} because they use a fixed sequence of transformation passes. SEER overcomes the phase-ordering challenge using e-graphs, customizing the MLIR pass order for each input program and optimization objective.

%%%%%%%%%%%%%%%%%%%%%%%%%%%%%%%%%%%%%%%%%%%%%%%%%%%%%
% METHODOLOGY
%%%%%%%%%%%%%%%%%%%%%%%%%%%%%%%%%%%%%%%%%%%%%%%%%%%%%
\section{Methodology}
\label{sec:methodology}

In this section, we describe the proposed source-to-source super-optimization tool for HLS. First, we provide an overview of the proposed SEER toolflow. We then introduce a new intermediate language, named SeerLang, that provides the first interface between MLIR and the \egg~e-graph library. Next, we explain the rewriting rules included in SEER and how to explore these rewriting rules in the e-graph, to construct $R''$ in Section~\ref{sec:motivating_example}. Finally, we describe the cost functions used for representation extraction for determining $t'$ in Section~\ref{sec:motivating_example}.

%%%%%%%%%%%%%%%%%%%%%%%%%%%%%%%%%%%%%%%%%%%%%%%%%%%%%
% SEER OVERVIEW
%%%%%%%%%%%%%%%%%%%%%%%%%%%%%%%%%%%%%%%%%%%%%%%%%%%%%
\subsection{SEER Overview}
To maximize generality and avoid targeting a particular HLS tool, SEER performs source-to-source transformation on the input software program and generates an efficient representation for HLS tools. SEER accepts C, C++, SystemC code, and other software programming languages that can be translated to MLIR. Figure~\ref{fig:seer_overview} illustrates a high-level overview of the SEER tool flow for HLS super-optimization. 

\begin{enumerate}[label={{\textcolor{jcred}{\large \textcircled{\raisebox{-0.9pt}{\normalsize \arabic*}}}}}]
\item The input program in C/C++/SystemC is parsed by Polygeist~\cite{moses2021polygeist}, a C (and C++) front end for MLIR, translating the program into the MLIR {\tt affine} or {\tt scf} dialects. We implemented MLIR transformation passes for converting a subset of SystemC. 
\item The SEER front end translates the MLIR into a new intermediate language, SeerLang, which provides an interface between MLIR and the e-graph library, \egg~\cite{willsey2021egg}. SeerLang is described in Section~\ref{sec:seerlang}.
\item From SeerLang an initial e-graph is constructed in \egg, where each e-class contains a single e-node.
\item SEER provides a set of patterns to \egg, which are used to search for rewriting opportunities in the e-graph, a process known as e-matching.
\item Once a pattern in the e-graph is matched, a validity condition is checked and a new equivalent SeerLang expression is constructed. Section~\ref{sec:rewriting_rules} describes SEER's rewrites.
\item If the rewrite is valid, the new SeerLang expression is unioned into the e-graph, as shown in Figure~\ref{fig:egraph_example}. The e-graph continues to grow until reaching a user defined limit, or until no new equivalent representations can be found. Rewriting in SEER is explained in Section~\ref{sec:exploring_rewrites}.
\item From the final e-graph, an extraction is performed to obtain an efficient implementation based on control path and data path hardware cost functions. The details of these cost functions are explained in Section~\ref{sec:cost_function}. 
\item The extracted SeerLang expression is translated back to the MLIR {\tt affine} or {\tt scf} dialects by the SEER back end, such that we can exploit existing MLIR back ends. 
\item The generated MLIR is converted back to SystemC using emitC~\cite{mlir_emitc}, such that the optimized program can be parsed by HLS tools. We extended the C back end to emit SystemC programs.
\item The equivalence between the original and transformed programs is proven by a formal equivalence checking tool, VC Formal from Synopsys~\cite{hector}, at SystemC level. The equivalence check steps are explained in Section~\ref{sec:equivalence_check}
\end{enumerate}

%%%%%%%%%%%%%%%%%%%%%%%%%%%%%%%%%%%%%%%%%%%%%%%%%%%%%
% SEERLANG
%%%%%%%%%%%%%%%%%%%%%%%%%%%%%%%%%%%%%%%%%%%%%%%%%%%%%
\subsection{SEER Intermediate Representation}
\label{sec:seerlang}

A key challenge for enabling MLIR exploration via e-graph rewriting in \egg~is that these two frameworks do not share a common representation language, and re-implementing either would require significant engineering effort. We identified three potential solutions for orchestrating them in the same toolflow. First, we could keep each MLIR representation in memory but removing redundancy among these versions is challenging, making the memory size unscalable. Second, we could keep a single representation and pass traces for obtaining each new MLIR representation. This leads to unscalable compilation time for reproducing the required representation. Finally, we decided to propose a new language named SeerLang in \egg~for translation to and from MLIR. 

In \egg, users define an S-expression based language similar to Common Lisp \cite{Steele1990CommonLanguage} to represent expressions. 

\texttt{term ::= (operator [term] [term]\ldots [term])}

\noindent
This language format allows users to concisely express rewrites. We defined a domain-specific representation, called SeerLang, that provides an interface between MLIR and \egg. The semantics of SeerLang are similar to MLIR as the representation is for translation only. SeerLang supports a subset of MLIR operations including operations in the {\tt affine}, {\tt scf}, {\tt memref} and {\tt arith} dialects, but can be extended to support other MLIR operations. In addition, SeerLang supports a {\tt seq} operator to encode the original program ordering between two operations.

\begin{figure}
\centering
\begin{tabular}{cb{0.48\textwidth}}
\begin{tabular}{b{0.16\textwidth} b{0.25\textwidth}}
\begin{lstlisting}[language=C, escapeinside={(*@}{@*)}, caption=C source, label=list:4, linewidth = 0.15\textwidth]
int x[8];
int i;
for (i=0;i<8;i++)
{
  int a = x[i];
  int b = a*2+1;
  x[i] = b;
}
\end{lstlisting}
&
\begin{lstlisting}[language=mlir, escapeinside={(*@}{@*)}, caption=MLIR code, label=list:5, linewidth = 0.25\textwidth]
affine.for %i=0 to 8 step 1 {
  %a=affine.load %x[%i] : memref<8xi32>
  %b0=arith.muli %a,2 : i32
  %b1=arith.addi %b0,1 : i32
  affine.store %b1, %x[%i] : memref<8xi32>
}
\end{lstlisting} 
\end{tabular} \\
% \vspace{-2em}
\begin{lstlisting}[language=seerlang, escapeinside={(*@}{@*)}, caption=SeerLang expression, label=list:6, linewidth = 0.42\textwidth]
(affine.for "affine.for_0" %i 0 8 1 none none none 
(block 
(seq
(affine.load i32 "%a" i32 (8) "%x" (%i))
(affine.store 
i32 (+ i32 i32 (* i32 i32 "%a" i32 2) i32 1) 
i32 (8) "%x" (%i))
)))
\end{lstlisting}
\end{tabular}
\caption{Example of SeerLang for expressing a {\tt for} loop and memory operations.}
\label{fig:seerlang_example}
\end{figure}

Here we introduce two key constructs in SeerLang, operation and block, inspired by MLIR. An operation takes a set of inputs and produces a set of results. An operation could be a data path operation like an {\tt add} operation or a {\tt mul} operation, or a control path operation like a function, a loop or an {\tt if} statement. A block contains a set of operations. In each block, the SeerLang front end analyses the data dependence between operations in the same block and reconstructs expression trees. A {\tt seq} operation is purely an annotation, preserving the original program order by keeping memory operations in the block and associating them using {\tt seq} operations. This facilitates memory dependence analysis for the transformation pass.

An example of SeerLang is shown in Figure~\ref{fig:seerlang_example}. Listing~\ref{list:4} shows a for loop that contains two memory operations. This is translated into Listing~\ref{list:5} in the MLIR {\tt affine} dialect. The {\tt for} loop in C is translated into an {\tt affine.for} operation because the loop contains only affine memory accesses. The memory operations are translated into {\tt affine} memory operations as the array index is a simple loop iterator and is in affine form. The arithmetic operations are translated to operations in the MLIR {\tt arith} dialect. The equivalent SeerLang of Listing~\ref{list:5} is shown in Listing~\ref{list:6}.

\begin{table}
\centering
\caption{Example SEER rewriting rules implemented directly in \egg. SEER contains 106 data path and gate-level rewrites~\cite{coward2022rover}. All datapath rewrites are signage and bitwidth dependent.}
\label{tab:rewriting_rules_mlir}
\resizebox{0.48\textwidth}{!}{%
\begin{tabular}{lccc}
\toprule
\cellcolor[HTML]{FFFFFF}Class & Pattern & Transformation \\
\midrule
Control Path & $(\texttt{seq}\;\;a\;\;(\texttt{seq}\;\; b\;\; c))$ & $(\texttt{seq}\;\; (\texttt{seq}\;\; a\;\; b)\;\; c)$ \\
 \midrule
 & \cellcolor[HTML]{DEEBF7}$(a\times b)\ll c$ & \cellcolor[HTML]{DEEBF7}$(a\ll c) \times b$ \\
  & $a \;?\; (b + c) : (d + e)$ & $(a \;?\; b : d) + (a \;?\; c : e)$ \\
  & \cellcolor[HTML]{DEEBF7}$(a\times b) + a$ & \cellcolor[HTML]{DEEBF7}$a\times (b+1)$ \\
  & $a\ll c$ & $a\times 2^c$ \\
  & \cellcolor[HTML]{DEEBF7}$(a\ll b)\ll c$ & \cellcolor[HTML]{DEEBF7}$a\ll(b+c)$ \\
 \multirow{-6}{*}{Data Path} & $-a$ & $\overline{a} + 1$ \\
 \midrule
  & $(a\&b)\oplus(a\&c)$ & $a\&(b\oplus c)$ \\
& \cellcolor[HTML]{DEEBF7}$a\oplus a$ & \cellcolor[HTML]{DEEBF7}$0$ \\
 \multirow{-3}{*}{Gate Level} & $\overline{a\& b}$ & $\overline{a} \| \overline{b}$ \\
\bottomrule
\end{tabular}
}
\end{table}

Translating into SeerLang from MLIR is nearly lossless since each operation in SeerLang keeps the type and operand information, except for the program order of independent data path operations. Independent operations are parallelized in hardware and its original program does not affect correctness. The data dependence is analyzed by the front end of SeerLang when translating from MLIR. For example, in Listing~\ref{list:6}, the arithmetic operations for {\tt \%b0} and {\tt \%b1} are converted into a nested expression at line 6. This recovers the data flow graph of the block for data path optimization. The operations with a potential data dependence are connected using {\tt seq} operations. In SEER, we assume there exists a data dependence between every two memory operations for simplicity. The memory operations are connected using {\tt seq} operations, such as the load and store operations in Listing~\ref{list:6}. This preserves the program order of memory access and ensures the correctness of memory transformation.

\begin{figure*}[t]
    \centering
    \includegraphics[width=\textwidth]{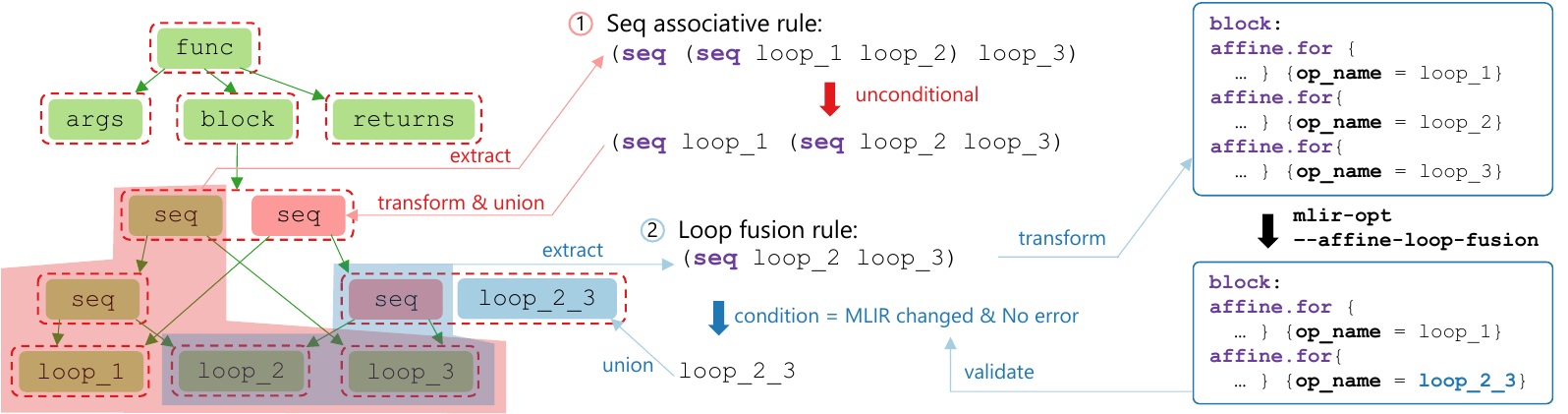}
    \caption{E-graph exploration of the motivational example (Figure~\ref{fig:motivating_example}) using SEER. The e-graph is simplified by merging subgraphs of loops into single nodes. The green nodes represents the initial e-graph obtained from Listing~\ref{list:1}. 
    \textcolor{jcredl}{\textcircled{\raisebox{-0.9pt}{\color{black}{1}}}} 
    illustrates an example of an unconditional rewrite for a sequential association inside \egg.
    For rewrite \textcolor{jcredl}{\textcircled{\raisebox{-0.9pt}{\color{black}{1}}}}, 
    the original sub-expression in the shaded red region is rewritten to the red node in the same e-class.
    \textcolor{jcbluel}{\textcircled{\raisebox{-0.9pt}{\color{black}{2}}}} illustrates an example of a conditional rewrite for loop fusion through MLIR. 
    For rewrite \textcolor{jcbluel}{\textcircled{\raisebox{-0.9pt}{\color{black}{2}}}}, 
    the original sub-expression in the shaded blue region is rewritten to the blue node in the same e-class.}
    \label{fig:example_rewrites}
\end{figure*}

\begin{figure*}[t]
    \centering
    \begin{minipage}[b][][b]{.55\textwidth}
        \centering
        \includegraphics[scale=0.57]{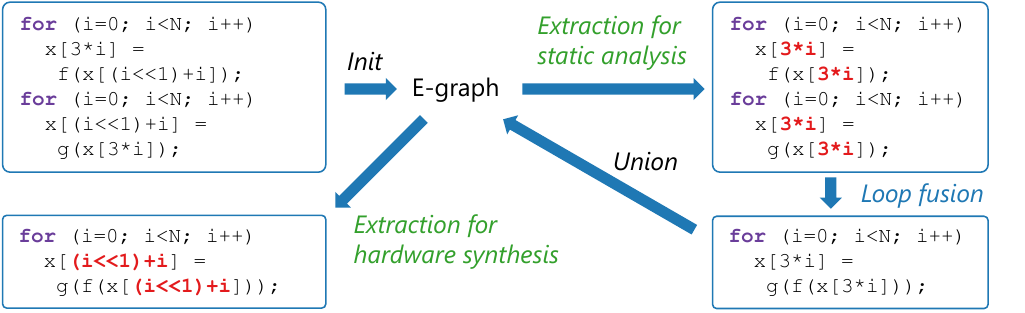}
    \caption{An example of extracting representations using cost functions \\ for static analysis and hardware synthesis.}
    \label{fig:static_analysis_cost}
    \end{minipage}%
    \begin{minipage}[b][][b]{0.45\textwidth}
        \centering
        \includegraphics[scale=0.57]{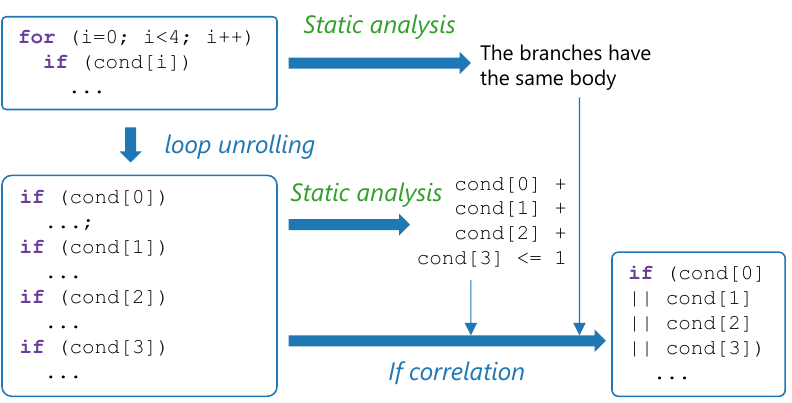}
    \caption{An example of using program invariants from static analysis one representation for transformation of another.}
    \label{fig:capability_enabling}
    \end{minipage}
\end{figure*}

%%%%%%%%%%%%%%%%%%%%%%%%%%%%%%%%%%%%%%%%%%%%%%%%%%%%%
% REWRITES
%%%%%%%%%%%%%%%%%%%%%%%%%%%%%%%%%%%%%%%%%%%%%%%%%%%%%
\subsection{Rewriting Rules}
\label{sec:rewriting_rules}

The rewriting rules in SEER enable the exploration of equivalent implementations of a program. SEER supports both internal rewrites, expressed directly in SeerLang, and external rewrites, expressed as MLIR passes. For internal rules, \egg~can directly apply them to add equivalent sub-expressions to the e-graph. A subset of these rules is shown in Table~\ref{tab:rewriting_rules_mlir}. External rules are implemented as dynamic rewrites in \egg, where we match against a SeerLang pattern and then construct an equivalent implementation using an external pass. In this construction, SeerLang must be translated into a compatible representation, modified by the external pass, and translated back to SeerLang. At this point \egg~can union this new sub-expression into the appropriate e-class. Such an approach makes it simple to implement new rules and enables the reuse of existing rules from other toolflows. SEER rewrites at different granularities, allowing it to simultaneously optimize at the control path level, data path level and gate level. 

First, the control path-level rewrites modify the control flow graph (CFG) of the original program. Particularly, we focus on the transformation of {\tt for} loops and {\tt if} statements. This includes ten MLIR passes for loop re-ordering, loop merging and if conversions. We maximize the reuse of available MLIR passes in upstream in SEER.

Most of the loop transformation passes are directly adopted from the MLIR/LLVM upstream and applied to SEER. The {\em loop unroll} pass performs complete unroll of a loop. This enables potential loop body reduction by other passes such as the memory forward pass. We disable exploring loop unrolling with different unrolling factors by default to improve scalability. It is provided as a user option. The {\em loop fusion}, {\em loop interchange} and {\em loop flatten} passes are existing compiler transformations which are directly mapped to SEER. The {\em loop perfection} pass converts a loop nest that contains code in its outer loop body and outside its inner loop body to a perfect loop nest. This is done by moving the code outside the inner loop body into the inner loop body with predicates. Loop perfection opens up opportunities for more loop transformation, such as loop interchanging and loop flattening.

The {\em if conversion} pass is used to convert {\tt if} statements to {\tt select} operations, reducing the control flow complexity. This has been widely used in the HLS code transformation for maximizing a data path region for pipelining. The {\em memory forward} pass removes redundant load and store operations in the code to reduce memory accesses. The upstream pass only removes store operations. We extend it to remove redundant load operations as well.

Customized MLIR passes can also be easily extended to SEER with the same interface. For instance, the {\em if correlation} pass is a customized pass which detects correlation among conditions of several sequential {\tt if} statements and merges them if the conditions are identical or disjoint. An example of {\tt if} correlation is described in Section~\ref{sec:analysis_capability_enabling}. The {\em memory reuse} pass moves a read-only memory access outside the loop. The {\em control flow mux} pass moves an operation in both branches of an {\tt if} statement outside the {\tt if} statement and select its args in the branches for resource sharing at the source level.

The data path-level rewrites modify the program at a finer grain and are mostly re-used from the e-graph based ROVER tool~\cite{coward2022rover,coward2023automating}. They include expression balancing, constant folding and manipulation, and strength reduction. Data path optimization is currently under-explored in existing commercial synthesis tools and recent synthesis-aware data path rewriting has been shown to reduce circuit area~\cite{coward2022rover,coward2023automating}. Two rewrites from Table~\ref{tab:rewriting_rules_mlir} are applied to the e-graph in Figure~\ref{fig:egraph_example}.

Finally, the gate-level rewrites also modify the program at the operator level but target bit-level hardware customization. Most gate-level rewrites are well exploited by the logic synthesis optimization in HLS tools. However, data path and bit-level rewriting can often interact providing a mutual benefit. SEER restricts the number of gate minimization techniques to improve scalability. We group these into the data path set.

%There are also existing hardware-specific rewriting rules for logic and algorithmic operations in an \egg-based tool named ROVER~\cite{coward2022rover}. These rules can be directly integrated into SEER because they both speak SeerLang. The integration of several rewrite sources highlights the extensibility of SEER.

%%%%%%%%%%%%%%%%%%%%%%%%%%%%%%%%%%%%%%%%%%%%%%%%%%%%%
% E-GRAPH REWRITING
%%%%%%%%%%%%%%%%%%%%%%%%%%%%%%%%%%%%%%%%%%%%%%%%%%%%%

\subsection{E-graph Rewriting for Super-optimization}
\label{sec:exploring_rewrites}
SEER alternates between iterations of control flow rewriting and data path rewriting.
% , as shown in Figure~\ref{fig:search_flow}. 
At each iteration all rules within the given rewrite set are applied, growing the e-graph. SEER interleaves the exploration of these rewrite sets since one might introduce more rewriting opportunities for the other. For instance, dead code elimination, a data path rewrite, can change the dependence constraints, enabling more control path rewrites. Loop fusion, a control path rewrite, can enable further rewrites for the fused loop body. 

Figure~\ref{fig:example_rewrites} shows the e-graph exploration of the motivating example introduced in Figure~\ref{fig:motivating_example}. To the initial e-graph, represented by the green nodes, SEER applies an internal rewrite rule from Table~\ref{tab:rewriting_rules_mlir}, seq associativity. The general rule, shown in the top middle of the figure, matches the sub-expression covered by the red shading and returns an equivalent SeerLang expression. This new expression is unioned into the matched e-class, where the new nodes are shown in red.  

Next SEER applies the external loop fusion MLIR transformation, that represents the transformation from Listing~\ref{list:1} to Listing~\ref{list:3}. The loop fusion rule searches for two sequential loops and checks if they satisfy the particular dependency constraints. First, the sub-expression covered by the blue shading matches the pattern of the loop fusion rewrite. The matched SeerLang is translated into the equivalent MLIR. Then SEER calls the existing loop fusion pass in MLIR, generating a new MLIR implementation. The loop fusion pass performs the dependence check internally before the transformation. If the dependence constraints are unsatisfied or the transformation fails, the pass returns the original MLIR. The loop fusion rule in \egg~checks that there were no errors in the pass and that the returned MLIR differs from the input then converts this back to SeerLang and performs the union. In this example, the result from fusing {\tt loop\_2} 
 and {\tt loop\_3} is added to the e-graph, as the blue {\tt loop\_2\_3} node.

Many other SEER rewrites can be applied to grow a larger e-graph than the one presented in Figure~\ref{fig:example_rewrites}. Thanks to the e-graph representation, despite fusing {\tt loop\_2} and {\tt loop\_3}, the fusion of {\tt loop\_1} and {\tt loop\_2} can also still be triggered. This would not be possible in a traditional compiler. Note that the fusion of {\tt loop\_1} and {\tt loop\_2\_3} will be attempted but will fail due to the validity checks. The e-graph grown after several rewriting iterations, represents the explored design space of equivalent implementations. From this e-graph SEER must now select an efficient HLS implementation.

% \begin{figure}
%     \centering
%     \includegraphics[width=0.35\textwidth]{imgs/seer_search_flow.pdf}
%     \caption{Exploration flow for rewriting rules.}
%     \label{fig:search_flow}
% \end{figure}

%%%%%%%%%%%%%%%%%%%%%%%%%%%%%%%%%%%%%%%%%%%%%%%%%%%%%
% ANALYSIS
%%%%%%%%%%%%%%%%%%%%%%%%%%%%%%%%%%%%%%%%%%%%%%%%%%%%%
% \subsection{Program Analysis \& Furthering Exploration}
\subsection{Deeper Optimization Opportunities}
\label{sec:analysis_capability_enabling}
Prior work observed how retaining multiple representations in an e-graph can improve program analysis~\cite{coward2023combining}. Here we observe a practical benefit of this, allowing SEER to discover implementations that are unreachable with existing compiler passes. SEER can learn program invariants from one representation which it can use to rewrite any equivalent representation.

% During the exploration, we observed that SEER improves the capability of existing static analysis techniques by providing a simpler and equivalent representation of the program. In addition, the program invariants extracted from a representation of the program can be used for the transformation of another equivalent representation. This results in a new representation that is not reachable by existing transformation passes.

Firstly, we shall describe how SEER can resolve loop dependence analysis limitations. Existing loop optimization passes use polyhedral analysis to detect any loop dependency issues. Such tools are unable to analyze memory access patterns that are not obviously affine. This introduces a tension, as a representation for efficient hardware synthesis could be complex for static analysis. For example in Figure~\ref{fig:static_analysis_cost}, the memory access index {\tt (i<<1)+i} is area-efficient in hardware because the shift operation is area-free in ASIC design, and only one adder is used. Polyhedral analysis tools will fail to interpret such a non-affine access pattern and conservatively may prevent subsequent loop optimizations.

Applying the data path rewrites to an e-graph containing {\tt (i<<1)+i} as shown in Figure~\ref{fig:egraph_example}, SEER discovers the equivalent \emph{affine} expression {\tt 3*i}, which is interpretable by the static analyzer. Such an expression may not be area-efficient since it uses a multiplier.

% In Figure~\ref{fig:static_analysis_cost}, a program of two sequential loops is being explored for rewriting, which contains array index expressions using shift and multiplication. Traditional compiler is capable to analyze memory accesses to {\tt 3*i+3} but struggles with the memory accesses to {\tt (i<<1)+i+3}. As a result, the static analyzer cannot determine the absence of memory dependence between the loop and fails fusing two loops.

Starting from an initial e-graph containing the input program in Figure~\ref{fig:static_analysis_cost}, SEER applies data path rewrites, discovering a representation where both memory indices are in affine form (top right). When SEER calls its loop fusion pass, it is presented with a choice of many equivalent loop implementations which it could pass to the external compiler pass. SEER aims to pass on analysis-friendly implementations, namely those with affine memory accesses. To achieve this SEER includes an analysis-friendly cost function, that assigns low cost to multiplications and additions, making affine expressions lower cost than alternative logic expressions. When a loop optimization triggers, SEER runs an extraction process on the matched e-class, using the analysis-friendly cost function, extracting affine memory access patterns where possible. In Figure~\ref{fig:static_analysis_cost}, SEER can successfully fuse the two loops (bottom right) and generate a final HLS program using the hardware-efficient but non-affine memory access pattern (bottom left). 

% The static analyzer can directly observe that the indices are the same and verify the absence of dependence between these loops. As a result, we can fuse these loops, leading to a representation shown on the bottom right. The representation is then unioned into the e-graph and continues being explored by other rewrites. For the final hardware design, SEER extracts another expression using the cost function for efficient hardware synthesis, as shown on the bottom left. In this representation, the loops are fused for better performance, and the array indices are expressed in an area-efficient way instead of affine forms. 

% The cost function for extracting static analysis-friendly representation is simpler. First, it sets the required operations in the pattern to zero cost, such as {\tt loop\_2} and {\tt loop\_3} when attempting to fuse them in Figure~\ref{fig:example_rewrites}. Second, we add heuristic-based weights onto the operations, particularly low weights to multiplications and additions, in order to make affine expressions lower cost than other expressions. The cost of each loop is the sum of the costs of the operations in its loop body, and the cost of each if statement is the sum of the costs of the operations in both branches.

The advantage is also seen in other programming constructs. The code in the top left of Figure~\ref{fig:capability_enabling} shows an {\tt if} statement in a {\tt for} loop. A possible transformation is loop unrolling, leading to straight-line code with four {\tt if} statements (bottom left). Assuming that all conditions are independent and at most one of them is true, it is possible to merge the conditions into a single block. However, existing compiler passes struggle to compare the source of the true branches for the {\tt if} statements, particularly when the code size is large. Fortunately, these invariants can be easily obtained from the original representation (top left), which is still present in the e-graph. With the invariants in place, the transformation is enabled and successful. This highlights another advantage of maintaining multiple representations in the e-graph.

%%%%%%%%%%%%%%%%%%%%%%%%%%%%%%%%%%%%%%%%%%%%%%%%%%%%%
% COST FUNCTION and EXTRACTION
%%%%%%%%%%%%%%%%%%%%%%%%%%%%%%%%%%%%%%%%%%%%%%%%%%%%%
\subsection{Cost Function Specifications}
\label{sec:cost_function}
SEER generates an efficient HLS implementation from the e-graph via a process, known as extraction. SEER combines a pair of theoretical cost functions to rank the equivalent implementations. We separate the extraction into a two-phase problem, first extract the control flow nodes that maximizes performance, then from the fixed control flow minimize the data path circuit area. 

The control path is usually parallelized, such that the latency of each data path is hidden by the pipeline. Pipelining is usually beneficial because it improves performance at a slightly higher area cost. SEER assumes all the loops are pipelined by default to achieve better performance.

\begin{figure}
    \centering
    \includegraphics[width=0.4\textwidth]{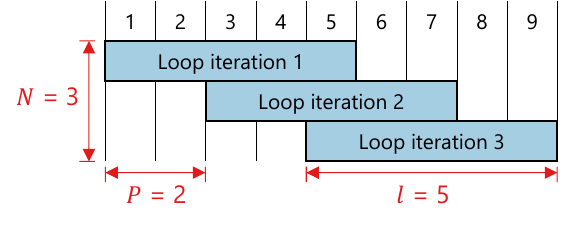}
    \caption{A schedule of a pipelined loop in HLS}
    \label{fig:pipelined_loop}
\end{figure}

The control flow cost function evaluates the latency of pipelined loops in terms of clock cycles. A pipelined loop has three scheduling constraints: the initiation interval $P$, iteration latency $l$, and loop trip count $N$. Figure~\ref{fig:pipelined_loop} shows an example of a simple pipelined loop. The initiation interval is the number of clock cycles between two consecutive loop iterations. The iteration latency is the latency of a single iteration in clock cycles. For this example, $P$ = 2 and $l$ = 5. These are typically constants because most HLS tools use static scheduling~\cite{zhang2013sdc, canis2014modulo}. The loop trip count represents the number of iterations. For this example, $N$ = 3. The total latency $L$ of a pipelined loop can be obtained based on the formula shown in Constraint~\ref{eqn:static_scheduling}~\cite{canis2014modulo}. SEER obtains the scheduling constraints of each loops in the original representation of the program by calling the HLS tool to schedule the original representation. 
\begin{align}
    L = (N-1) \times P + l \label{eqn:static_scheduling}
\end{align}
In order to improve scalability, we approximate the schedule constraints of the newly generated loops during the exploration from the existing scheduling constraints of the original loops. This approximation facilitates exploration at scale, avoiding calls to the HLS scheduler for each new representation. 

For each loop in the initial representation, SEER obtains $(P, l, N, A)$ from the initial HLS run, where $A$ is the set of memory accesses in the loop. $A$ is a resource constraint used for estimating the upper bound of throughput based on the memory bandwidth at run time. For instance, each loop in Listing~\ref{list:1} has $|A|$ = 2.

Here we provide three key examples of loop transformations at different levels, loop fusion, loop flattening and loop unrolling. First, let $(P_1, l_1, N_1, A_1)$ and $(P_2, l_2, N_2, A_2)$ be the scheduling constraints for two sequential loops to be fused. Let $M(A)$ be the maximum number of memory accesses to a single array, and assume all the BRAM blocks are single-port. The fused loop will have constraints $(P', l', N', A')$ as follows. 
\begin{align*}
    l' &= max(l_0, l_1) &N' &= max(N_0, N_1)  \\
    A' &= A_0 \cup A_1  & P' &= max(P_1, P_2, M(A'))  
\end{align*}
Second, if the outer loop and the inner loop of a perfect loop nest have scheduling constraints $(P_{outer}, l_{outer}, N_{outer}, A_{outer})$ and $(P_{inner}, l_{inner}, N_{inner}, A_{inner})$ respectively, the flattened loop then has scheduling constraints $(P_{inner}, l_{inner}, N_{inner} \times N_{outer}, A_{inner})$. Finally, if a loop with scheduling constraints $(P, l, N, A)$ is unrolled, the scheduling constraints after the transformation are $(1, N \times l, 1, N \times A)$. 

To formulate the extraction, let $E$ denote the set of all program representations in the e-graph. SEER assigns a latency, $L(n)$, to each e-node, $n$:
 \begin{equation}
    L(n)=
    \begin{cases}
      (N_n-1) \times P_n + l_n, & \text{if}\ n \text{ is a loop} \\
      0, & \text{otherwise}
    \end{cases}
  \end{equation}
, where we denote the scheduling constraints of a loop node $n$ by $(P_n, l_n, N_n, A_n)$. A completely unrolled loop is still considered a loop with an iteration count of 1 to avoid zero cost during the extraction. The if statements are extracted based on the data path cost function.
The objective for control flow extraction is then to minimize the sum of loop latencies. 
\begin{equation}
\begin{aligned}
\min_{e} \quad & \sum_{n \in \textrm{e}} L(n)\\
\textrm{s.t.} \quad & e \in E\\
\end{aligned}
\end{equation}

We use a greedy extraction method for control flow, selecting the lowest cost loop implementation in each control flow e-class. Since data path nodes are considered zero cost at this stage, control flow extraction returns a subset of representations $E'\subseteq E$, that all share the same optimized control flow. SEER must now select an efficient data path implementation from these remaining implementations. 

For the data path, SEER minimizes area rather than latency, as operation latencies are often hidden by loop pipelining. SEER leverages an existing cost function from ROVER to extract the minimal circuit area expression in each block~\cite{coward2022rover}. ROVER assigns an area cost, $A(n)$, to each e-node, $n$, based on a bitwidth-dependent gate count. 
%Let $N$ be the set of operations in a program representation. Previous control path extraction returns a $F' \subseteq N$ as the extracted loops. 
The objective for data path extraction is then:
% \begin{align}
    % \min_{\textrm{e}\in E'} \quad \sum_{n \in \textrm{e}} A(n)
% \end{align}
\begin{equation}
\begin{aligned}
\min_{e} \quad & \sum_{n \in \textrm{e}} A(n)\\
\textrm{s.t.} \quad & e \in E'\\
\end{aligned}
\end{equation}
In ROVER, data path extraction is formulated as an integer linear programming (ILP) problem~\cite{Wang2020SPORES:Algebra,coward2022rover}, solved using the Coin-Or CBC solver~\cite{forrest2005cbc}. The ILP returns a single SeerLang representation, which is passed to the SEER back end.

\subsection{Verification}
\label{sec:equivalence_check}

Our work benefits from SEER orchestrating existing transformation passes for super-optimization exploration. However, these passes often need to be verified. We adopt a translation validation approach based on the \egg~proof production feature~\cite{Flatt2022SmallClosurec}. SEER traces back the intermediate forms to the original program from the extracted representation, generating SystemC for each step. It then generates a sequence of equivalence checks, constructing a sound chain of reasoning that the original and generated programs are functionally equivalent. We prove the equivalence checks using the Synopsys VC Formal equivalence checker.  
%The verification time is negligible compared to the hardware synthesis time.

%Verification is crucial to ensure the optimized representation of the program must be functionally equivalent to the original representation.

% To ensure soundness, we use formal verification for equivalence checks on these representations. SEER extracts the sequence of rewrite steps and translates the intermediate representations between every two steps in SeerLang to SystemC. Then it uses Synopsis HECTOR to check the equivalence between the representations at each step. Since each transformation step has small changes, the verification time is negligible compared to the hardware synthesis time.

\definecolor{areacolor}{HTML}{d7191c}
\definecolor{timecolor}{HTML}{2b83ba}
\definecolor{powercolor}{HTML}{fdae61}
\definecolor{ppacolor}{HTML}{abdda4}

\begin{figure*}[t]
\centering
\begin{minipage}[b]{0.3\textwidth}
\begin{lstlisting}[language=C, escapeinside={(*@}{@*)}, linewidth = 0.9\textwidth]
for (j=1; j<4; j++) {
  for (i=0; i<4; i++) { (*@\textcolor{jcredl}{\textcircled{\raisebox{-0.9pt}{\bf 1}}}@*)
    if (list[j][i] 
     && enable[i](*@\textcolor{jcredl}{\textcircled{\raisebox{-0.9pt}{\bf 2}}}@*))(*@\textcolor{jcredl}{\textcircled{\raisebox{-0.9pt}{\bf 3}}}@*)
    { (*@\textcolor{jcredl}{\textcircled{\raisebox{-0.9pt}{\bf 4}}}@*) 
      sum++; 
      enable[0] = false;
      enable[1] = false;
      enable[2] = false;
      ennable[3] = false; (*@\textcolor{jcredl}{\textcircled{\raisebox{-0.9pt}{\bf 6}}}@*) 
    }
    if (list[j][i]) (*@\textcolor{jcredl}{\textcircled{\raisebox{-0.9pt}{\bf 5}}}@*)
      enable[i] = true;
  }
}
\end{lstlisting}
\caption{{\tt byte\_enable\_calc}}
\label{fig:case_study_1}
\end{minipage}
\begin{minipage}[b]{0.68\textwidth}
\begin{tikzpicture}
\pgfplotsset{every x tick label/.append style={font=\footnotesize}}
\begin{axis}[
    width=0.6\textwidth, height=60mm, 
    bar width=5,
    ybar=0pt, ymode=log,
    ylabel near ticks, 
    ymin=0.1, ymax=2,
    legend cell align={left},
    ylabel={Normalised to baseline - $\times$},
    xlabel={\tt byte\_enable\_calc},
    ytick={0.1, 0.5, 1, 2},
    log ticks with fixed point,
    major x tick style = transparent,
    symbolic x coords={1,3,4,5,6,7,8},
    legend columns=4,
    xticklabels={SEER (C), SEER, Manual, SEER (Manual)},
    xtick=data
    ]
\addplot[timecolor, draw=timecolor, fill=timecolor] table  
[x=x, y=time, col sep=space] {
x	area	time	power	ppa
1	0.742268041	0.757758621	0.969387755	0.549090909
3	0.701030928	0.207758621	1.040816327	0.151818182
4	0.68556701	0.300862069	1.126530612	0.232727273
5	0.740341073	0.285714286	0.891805746	0.189337209
};
\addplot[areacolor, draw=areacolor, fill=areacolor] table  
[x=x, y=area, col sep=space] {
x	area	time	power	ppa
1	0.742268041	0.757758621	0.969387755	0.549090909
3	0.701030928	0.207758621	1.040816327	0.151818182
4	0.68556701	0.300862069	1.126530612	0.232727273
5	0.740341073	0.285714286	0.891805746	0.189337209
};
\addplot[powercolor, draw=powercolor, fill=powercolor] table  
[x=x, y=power, col sep=space] {
x	area	time	power	ppa
1	0.742268041	0.757758621	0.969387755	0.549090909
3	0.701030928	0.207758621	1.040816327	0.151818182
4	0.68556701	0.300862069	1.126530612	0.232727273
5	0.740341073	0.285714286	0.891805746	0.189337209
};
\addplot[ppacolor, draw=ppacolor, fill=ppacolor] table  
[x=x, y=ppa, col sep=space] {
x	area	time	power	ppa
1	0.742268041	0.757758621	0.969387755	0.549090909
3	0.701030928	0.207758621	1.040816327	0.151818182
4	0.68556701	0.300862069	1.126530612	0.232727273
5	0.740341073	0.285714286	0.891805746	0.189337209
};
\legend{\footnotesize Performance, \footnotesize Area, \footnotesize Power, \footnotesize PPA}
\end{axis}
\end{tikzpicture}
% \caption{Hardware results of {\tt byte\_enable\_calc} using different approaches.}
% \label{fig:case_study_bar_chart}
% \end{minipage}
% \begin{minipage}[b]{0.26\textwidth}
\begin{tikzpicture}
\pgfplotsset{every x tick label/.append style={font=\footnotesize}}
\begin{axis}[
    width=0.45\textwidth, height=60mm, bar width=5,
    ybar=0pt, ymode=log,
    enlarge x limits=0.3,
    ylabel near ticks, 
    ymin=0.1, ymax=2,
    xlabel={\tt seq\_loops},
    ytick={0.1, 0.5, 1, 2},
    log ticks with fixed point,
    major x tick style = transparent,
    symbolic x coords={so, ro, s},
    xticklabels={SEER (C), SEER (D), SEER},
    xtick=data
    ]
\addplot[timecolor, draw=timecolor, fill=timecolor] table  
[x=x, y=time, col sep=space] {
x	area	time	power	ppa
so	1.241739032	0.280356187	1.642010378	0.571631793
ro	0.905548664	1.066711141	1.197678268	1.15690792
s	1.367091463	0.126848319	1.066784468	0.184994566
};
\addplot[areacolor, draw=areacolor, fill=areacolor] table  
[x=x, y=area, col sep=space] {
x	area	time	power	ppa
so	1.241739032	0.280356187	1.642010378	0.571631793
ro	0.905548664	1.066711141	1.197678268	1.15690792
s	1.367091463	0.126848319	1.066784468	0.184994566
};
\addplot[powercolor, draw=powercolor, fill=powercolor] table  
[x=x, y=power, col sep=space] {
x	area	time	power	ppa
so	1.241739032	0.280356187	1.642010378	0.571631793
ro	0.905548664	1.066711141	1.197678268	1.15690792
s	1.367091463	0.126848319	1.066784468	0.184994566
};
\addplot[ppacolor, draw=ppacolor, fill=ppacolor] table  
[x=x, y=ppa, col sep=space] {
x	area	time	power	ppa
so	1.241739032	0.280356187	1.642010378	0.571631793
ro	0.905548664	1.066711141	1.197678268	1.15690792
s	1.367091463	0.126848319	1.066784468	0.184994566
};
% \legend{SEER-only, ROVER-only, SEER}
\end{axis}
\end{tikzpicture}
\caption{Normalized results of hardware designs for {\tt byte\_enable\_calc} and {\tt seq\_loops} using different approaches compared to the baseline results of hardware designs from original programs. The results by ROVER-only on {\tt byte\_enable\_calc} are the same as the baseline.}
\label{fig:case_study_bar_chart}
\end{minipage}
\end{figure*}

\section{Experiments}
\label{sec:experiments}

We evaluate SEER on a set of benchmarks. We compare SEER with vanilla Cadence Stratus HLS~\cite{stratus_hls}, a leading ASIC HLS tool, and the data path optimizer ROVER~\cite{coward2022rover}, as a hardware optimizer that also uses e-graphs. We do not compare against the related works~\cite{yehpca2022scalehls, zhao2022fpl} mentioned in Section~\ref{sec:mlir_background} since they target FPGAs and we target ASICs. 
%They still suffer from the phase-ordering problem in HLS. 

We assume that the designer has no hardware knowledge. To ensure fairness, we synthesize both the original, ROVER-generated, and SEER-generated programs using the same Stratus HLS configuration. We evaluate the impact of SEER on circuit area, performance in wall clock time and power. The total clock cycles were obtained from Stratus HLS co-simulation. The area and power results were obtained from the Post \& Route report from Stratus HLS. We use Stratus HLS version 22.02 and target the built-in 45nm technology library.
% All our designs are formally verified using the equivalence checker, Synopsys VC Formal, with the original programs \gc{Is this actually true?}.

\subsection{Benchmarks}

We combine artificially constructed, Intel provided and open-source benchmarks from the MachSuite~\cite{reagen2014machsuite} set. The benchmarks implement algorithms as low-level kernels suitable for hardware acceleration. We aim to evaluate SEER on super-optimization for 1) different application programs and 2) different implementations of the same application program using different algorithms. We use the following benchmarks:
\begin{description}[leftmargin=!,font={\ttfamily}] 
\item[seq\_loops] represents the sequential loop example shown in Figure~\ref{fig:static_analysis_cost}, amenable to loop fusion.
\item[byte\_enable\_calc] pre-processes and combines multiple messages into one. Widely used in computer architectures. 
% It combines as many messages as possible until it finds a clash in the resource being used (\textit{e.g.} a byte enable bit, cache bank or output port) and has been widely used in computer architectures. 
% \jc{Double check the text with Sam}.
\item[kmp] is an implementation of the Knuth-Morris-Pratt algorithm~\cite{kmp} for string matching.
\item[gemm (ncubed/blocked)] is a naive/blocked implementation of dense matrix multiplication. The ncubed algorithm is unoptimized and has a complexity of $O(n^3)$. The blocked algorithm~\cite{gemm_blocked} provides better locality.
\item[md (grid/knn)] simulates molecular dynamics using N-body methods to compute local forces. The grid implementation uses spatial decomposition from polyhedral transformations. The knn implementation was originally from the SHOC benchmark suite~\cite{md_knn} and uses K-nearest neighbors.
\item[sort (merge/radix)] sorts an integer array. Merge uses the merge sort algorithm~\cite{cole1988parallel}, and the radix implementation compares 4-bits blocks at a time.
\end{description}
{\tt seq\_loops} is made artificially to demonstrate a simple example, and {\tt byte\_enable\_calc} is production code provided by Intel. The rest of the benchmarks are directly obtained from the MachSuite~\cite{reagen2014machsuite} benchmark set.

\begin{table}[t]
\caption{Case study on the {\tt byte\_enable\_calc} benchmark. SEER (D) only explores data path optimizations and SEER (C) only explores control path optimizations. SEER (Manual) explores the manually optimized design. CP = Critical Path. ET = Execution Time. PPA = Performance Power Area product.}
\label{tab:case_study}
\centering
\resizebox{0.48\textwidth}{!}{%
\begin{tabular}{lrrrrrr}
\toprule
Approaches & \multicolumn{1}{c}{\makecell[c]{Area \\ ($\mu$m$^2$)}} & \multicolumn{1}{c}{Cycles} & \multicolumn{1}{c}{\makecell[c]{CP \\ (ns)}} & \multicolumn{1}{c}{\makecell[c]{ET \\ (ns)}} & \multicolumn{1}{c}{\makecell[c]{Power \\ (mW)}} & \multicolumn{1}{c}{\makecell[c]{PPA \\ ($yW \cdot m^2 \cdot s$)}} \\
\midrule
Baseline & 1.94 & 119 & 0.976 & 116 & 0.49 & 110 \\
SEER (D) & 1.94 & 119 & 0.976 & 116 & 0.49 & 110 \\
SEER (C) & 1.44 & 81 & 1.09 & 87.9 & 0.475 & 60.4 \\
SEER & 1.36 & \best{29} & \best{0.831} & \best{24.1} & 0.51 & \best{16.7} \\
Manual & \best{1.33} & 42 & \best{0.831} & 34.9 & 0.552 & 25.6 \\
SEER (Manual) & 1.44 & 34 & 0.976 & 33.2 & \best{0.44} & 20.8 \\
\bottomrule
\end{tabular}}
\end{table}
\begin{table*}
\centering
\caption{Evaluation of SEER over a set of benchmarks. Area in $\mu$m$^2$. Total Cycles in 1000's. Critical Path in ns. Power in mW.}
\label{tab:overall_results}
\resizebox{\textwidth}{!}{%
\begin{tabular}{lrrrrrrrrrrrr}
\toprule
\multirow{2}{*}{Benchmarks} & \multicolumn{4}{c}{Baseline} & \multicolumn{4}{c}{ROVER} & \multicolumn{4}{c}{SEER} \\
\cmidrule(lr){2-5}
\cmidrule(lr){6-9}
\cmidrule(lr){10-13}
 & \multicolumn{1}{c}{Area} & \multicolumn{1}{c}{Total Cycles} & \multicolumn{1}{c}{Critical Path} & \multicolumn{1}{c}{Power} & \multicolumn{1}{c}{Area} & \multicolumn{1}{c}{Total Cycles} & \multicolumn{1}{c}{Critical Path} & \multicolumn{1}{c}{Power} & \multicolumn{1}{c}{Area} & \multicolumn{1}{c}{Total Cycles} & \multicolumn{1}{c}{Critical Path} & \multicolumn{1}{c}{Power} \\
\midrule
{\tt seq\_loops} & 1.77 & 1.5 & 0.943 & {\color{jcgreen} \textbf{0.224}} & {\color{jcgreen} \textbf{1.61}} & 1.6 & 0.943 & 0.268 & 2.42 & {\color{jcgreen} \textbf{0.203}} & {\color{jcgreen} \textbf{0.883}} & 0.238 \\
{\tt kmp} & 8.09 & 357 & 1.53 & 0.581 & 8.09 & 357 & 1.53 & 0.581 & {\color{jcgreen} \textbf{7.72}} & {\color{jcgreen} \textbf{292}} & {\color{jcgreen} \textbf{1.42}} & {\color{jcgreen} \textbf{0.546}} \\
{\tt gemm (blocked)} & {\color{jcgreen} \textbf{14.4}} & 4620 & 1.16 & {\color{jcgreen} \textbf{0.735}} & {\color{jcgreen} \textbf{14.4}} & 4620 & 1.16 & {\color{jcgreen} \textbf{0.735}} & 16.3 & {\color{jcgreen} \textbf{537}} & {\color{jcgreen} \textbf{1.11}} & 3.44 \\
{\tt gemm (ncubed)} & {\color{jcgreen} \textbf{11.8}} & 3410 & 0.971 & {\color{jcgreen} \textbf{0.972}} & {\color{jcgreen} \textbf{11.8}} & 3410 & 0.971 & {\color{jcgreen} \textbf{0.972}} & 12.7 & {\color{jcgreen} \textbf{535}} & {\color[HTML]{333333} 0.971} & 5.83 \\
{\tt md (grid)} & {\color{jcgreen} \textbf{132}} & 1480 & 1.55 & {\color{jcgreen} \textbf{2.31}} & {\color{jcgreen} \textbf{132}} & 1480 & 1.55 & {\color{jcgreen} \textbf{2.31}} & 180 & {\color{jcgreen} \textbf{346}} & {\color{jcgreen} \textbf{1.54}} & 5.07 \\
{\tt md (knn) } & {\color{jcgreen} \textbf{107}} & 303 & 1.19 & 2.34 & {\color{jcgreen} \textbf{107}} & 303 & 1.19 & {\color[HTML]{333333} 2.27} & 127 & {\color{jcgreen} \textbf{8.25}} & {\color{jcgreen} \textbf{1.14}} & {\color{jcgreen} \textbf{1.62}} \\
{\tt sort (merge)} & 26.4 & 238 & 1.42 & 1.56 & 26.4 & 238 & 1.42 & 1.56 & {\color{jcgreen} \textbf{14.8}} & {\color{jcgreen} \textbf{153}} & {\color{jcgreen} \textbf{1.24}} & {\color{jcgreen} \textbf{1.36}} \\
{\tt sort (radix)} & 10.8 & 223 & 1.39 & 0.262 & 10.8 & 223 & 1.39 & 1.06 & {\color{jcgreen} \textbf{9.05}} & {\color{jcgreen} \textbf{136}} & {\color{jcgreen} \textbf{1.28}} & {\color{jcgreen} \textbf{1.02}} \\ 
\midrule
{\bf Norm. Geom. Mean.} & 1$\times$ & 1$\times$ & 1$\times$ & 1$\times$ & {\color{jcgreen} \textbf{0.99$\times$}} & 1.01$\times$ & 1$\times$ & {\color{jcgreen} \textbf{1.4$\times$}} & 1.06$\times$ & {\color{jcgreen} \textbf{0.34$\times$}} & {\color{jcgreen} \textbf{0.95$\times$}} & 2.54$\times$ \\
\bottomrule
\end{tabular}
}
\end{table*}
\definecolor{areacolor}{HTML}{d7191c}
\definecolor{timecolor}{HTML}{2b83ba}
\definecolor{powercolor}{HTML}{fdae61}
\definecolor{ppacolor}{HTML}{abdda4}

\begin{figure*}
\centering
\begin{minipage}[b][][b]{0.55\textwidth}
\begin{tikzpicture}
\pgfplotsset{every x tick label/.append style={font=\footnotesize}}
\begin{axis}[
    width=\textwidth, height=50mm, 
    bar width=5,
    ybar=0pt, ymode=log,
    ylabel near ticks, 
    ymin=0.01, ymax=10,
    xticklabel style={rotate=15, yshift=5pt, xshift=-4pt},
    legend cell align={left},
    ylabel={Normalised to baseline - $\times$},
    ytick={0.01, 0.1, 0.5, 1, 2, 5, 10},
    log ticks with fixed point,
    major x tick style = transparent,
    symbolic x coords={2,3,4,5,6,7,8},
    xticklabels={MLIR, ROVER, SEER},
    legend columns=4,
    xticklabels={{\tt kmp} ,{\tt gemm (blocked)} ,{\tt gemm (cubed)} ,{\tt md (grid)} ,{\tt md (knn) } ,{\tt sort (merge)} ,{\tt sort (radix)} },
    legend style={at={(0.03,0.17)},anchor=west},
    xtick=data
    ]
\addplot[timecolor, draw=timecolor, fill=timecolor] table  
[x=x, y=time, col sep=space] {
x	area	time	power	ppa
2	1.127237807	0.110568038	4.686029808	0.584050228
3	0.953971275	0.75956934	0.939592073	0.680835305
4	1.35864718	0.233061483	2.191324701	0.693879299
5	0.559958187	0.561422456	0.869031301	0.273200065
6	1.072409741	0.156664515	5.998775408	1.007845573
7	1.187810991	0.025959881	0.694955652	0.021429257
8	0.83550827	0.562987931	3.90528655	1.836972875
};
\addplot[areacolor, draw=areacolor, fill=areacolor] table  
[x=x, y=area, col sep=space] {
x	area	time	power	ppa
2	1.127237807	0.110568038	4.686029808	0.584050228
3	0.953971275	0.75956934	0.939592073	0.680835305
4	1.35864718	0.233061483	2.191324701	0.693879299
5	0.559958187	0.561422456	0.869031301	0.273200065
6	1.072409741	0.156664515	5.998775408	1.007845573
7	1.187810991	0.025959881	0.694955652	0.021429257
8	0.83550827	0.562987931	3.90528655	1.836972875
};
\addplot[powercolor, draw=powercolor, fill=powercolor] table  
[x=x, y=power, col sep=space] {
x	area	time	power	ppa
2	1.127237807	0.110568038	4.686029808	0.584050228
3	0.953971275	0.75956934	0.939592073	0.680835305
4	1.35864718	0.233061483	2.191324701	0.693879299
5	0.559958187	0.561422456	0.869031301	0.273200065
6	1.072409741	0.156664515	5.998775408	1.007845573
7	1.187810991	0.025959881	0.694955652	0.021429257
8	0.83550827	0.562987931	3.90528655	1.836972875

};
\addplot[ppacolor, draw=ppacolor, fill=ppacolor] table  
[x=x, y=ppa, col sep=space] {
x	area	time	power	ppa
2	1.127237807	0.110568038	4.686029808	0.584050228
3	0.953971275	0.75956934	0.939592073	0.680835305
4	1.35864718	0.233061483	2.191324701	0.693879299
5	0.559958187	0.561422456	0.869031301	0.273200065
6	1.072409741	0.156664515	5.998775408	1.007845573
7	1.187810991	0.025959881	0.694955652	0.021429257
8	0.83550827	0.562987931	3.90528655	1.836972875
};
\legend{\footnotesize Performance, \footnotesize Area, \footnotesize Power, \footnotesize PPA}
\end{axis}
\end{tikzpicture}
\caption{Normalized results for the SEER generated programs, across the key hardware efficiency metrics for the benchmarks evaluated.}
\label{fig:overall_bar_chart}
\end{minipage} 
\begin{minipage}[b][][b]{0.4\textwidth}
\begin{table}[H]
\centering
\caption{The size of the e-graphs and total search times for the evaluated benchmarks.}
\label{tab:egraph_size}
\resizebox{0.75\textwidth}{!}{%
\begin{tabular}{lrr}
\toprule
Benchmarks & \multicolumn{1}{c}{Nodes} & \multicolumn{1}{c}{Time - s} \\
\midrule
{\tt byte\_enable\_calc} & 31769 & 161 \\
{\tt seq\_loops} & 11906 & 0.3 \\
{\tt kmp} & 504 & 2.59 \\
{\tt gemm (blocked)} & 10947 & 20.2 \\
{\tt gemm (ncubed)} & 114 & 1.3 \\
{\tt md (grid)}& 44328 & 61.4 \\
{\tt md (knn)} & 43580 & 52.2 \\
{\tt sort (merge)} & 310 & 2.59 \\
{\tt sort (radix)} & 273 & 2.5 \\
\bottomrule
\end{tabular}}
\end{table}
\end{minipage}
\end{figure*}

\subsection{Intel Production Code Case Study}
\label{sec:intel_production_code}

We first provide a case study of applying SEER to a snippet of unoptimized Intel production code, where the source code is shown in Figure~\ref{fig:case_study_1}. The HLS tool cannot synthesize efficient hardware because of the data dependence on {\tt message} across loop iterations. The control path shown as {\tt if} statements are also unnecessary and are challenging for the tool to interpret. The following are potential optimization opportunities:

\begin{enumerate}[label={{\textcolor{jcred}{\large \textcircled{\raisebox{-0.9pt}{\normalsize \arabic*}}}}}]
    \item {\em Loop Unroll: } The iteration counts of both loops are small. It may be beneficial to unroll the loops for more data parallelism at low area overhead.
    \item {\em Memory Forward: } There are multiple load and store operations to {\tt message}. These operations could be folded to reduce data dependence on {\tt message}.
    \item {\em If Correlation: } When the loop is unrolled, the if conditions in different loop iterations may be correlated as shown in Figure~\ref{fig:capability_enabling}.
    \item {\em If Conversion: } The true branch of the {\tt if} statement at line 13 is a single line and could be rewritten as a multiplexer. 
    \item {\em Mux Reduction: } The same true branch updates a single bit using a constant, which could be directly fetched from the {\tt if} condition. The same applies to lines 7-10.
    \item {\em Gate Reduction: } The logic expressions in the loop body could be simplified once folded into a single data path using the transformation steps above.
\end{enumerate}

The results for the code in Figure~\ref{fig:case_study_1} are shown on the left of Figure~\ref{fig:case_study_bar_chart}. In the figure, smaller values indicate better results. In addition to the results obtained from the original program and SEER, we also obtained the results from the manually optimized design by Intel hardware experts. Furthermore, we gave the manually optimized design to SEER for further optimization. We made the following observations:

\begin{itemize}
    \item ROVER (SEER (D)) could not optimize the input program because the data paths are separated by control operations.
    \item Exploring MLIR passes (SEER (C)) improves performance and area, due to the conversion of control path operations to data path operations and memory forwarding.
    \item By combining ROVER rewrites and MLIR passes SEER achieves significantly better performance improvements and area reduction. The performance of SEER-generated hardware even outperforms the manually optimized hardware design with a small area overhead.
    \item SEER achieves the best hardware results among automatically generated designs, and SEER even optimizes the manually optimized hardware design by hardware experts, pushing the limit in both performance and area.
\end{itemize}
The detailed results for {\tt byte\_enable\_calc} are shown in Table~\ref{tab:case_study}. We also observed similar results for benchmark {\tt seq\_loops}, in the left of Figure~\ref{fig:case_study_bar_chart}:

\begin{itemize}
    \item Exploring ROVER optimization only (SEER (D)) achieves area reduction in the data path.
    \item Exploring MLIR passes only (SEER (C)) achieves performance improvements with area and power overhead.
    \item Exploring both ROVER and MLIR passes with SEER achieves the best performance with less power overhead due to the interaction between ROVER rewrites and MLIR passes, which enables more optimization opportunities.
\end{itemize}

\subsection{Overall Results}

The improvements on other benchmarks are shown in Figure~\ref{fig:overall_bar_chart}, and the detailed results are shown in Table~\ref{tab:overall_results}. SEER has achieved better performance for all the benchmarks by enabling automatic loop pipelining. Stratus HLS cannot auto-pipeline loops without human guidance. Loop pipelining also causes additional area and power overhead. Overall, the PPA for most benchmarks is improved. In the case of the {\tt sort (radix)} benchmark, the loop has a small trip count, so loop pipelining only provides marginal performance improvements at significant power overhead. Since our cost functions do not consider power, SEER only focuses on performance and area. This potentially introduces inefficient power in the final hardware design. On average, SEER achieved a speedup of 2.9$\times$ with 0.06$\times$ area overhead, and a PPA of 8.9$\times$ the original design.
The size of e-graph and exploration time for each benchmark are shown in Table~\ref{tab:egraph_size}. 
%The exploration times are negligible compared to the hardware synthesis time.

\section{Conclusion}
\label{sec:conclusion}

This paper described an approach to resolving the phase-ordering problem for HLS. By simultaneously exploring optimizations at different granularities in an e-graph, our approach opens up a larger optimization space for an arbitrary program than existing HLS works. We demonstrated how high-level control optimizations and low-level data path optimizations can mutually benefit, enabling further optimization opportunities. We model the hardware performance for the control path at a software abstraction level and determine efficient HLS code for high throughput and area efficiency. 

We implemented a toolflow, SEER, that uses an e-graph to orchestrate high-level software optimizations in MLIR and low-level hardware optimizations in ROVER. We introduced a new intermediate language, SeerLang, that interfaces the \egg~library and MLIR. We evaluated SEER on open-source benchmarks and an Intel-provided case study, demonstrating an average speedup of 2.9$\times$ with minimal area overhead. 

Our future work will involve several directions. First, from the programming language point of view, we plan to improve SeerLang for deeper integration with MLIR and \egg~for efficient translation. This would also resolve some engineering challenges since most MLIR passes target entire functions rather than local transformations. Second, we plan to extend the exploration space and granularities by integrating optimization techniques from other MLIR projects, such as CIRCT~\cite{circt} and POLSCA~\cite{zhao2022fpl}. We will investigate parallel e-graph exploration using multiple threads to improve scalability. Further efficiency gains could be made by partitioning the e-graph and exploring different sub-graphs independently. Lastly, we will evaluate SEER on larger benchmarks to understand the practical limitations of the approach.

% Future work:
% \begin{enumerate}
%     \item Better IR - remove the need to transform between \egg \& MLIR
%     \item Orchestration for efficiency - localized optimization initially
%     \item MLIR passes target functions - practical issue - can we locally target transforms in the e-graph
% \end{enumerate}

\bibliographystyle{plain}
\bibliography{ref}

\end{document}